\documentstyle[prb,aps,epsf]{revtex}

\begin{document}
\twocolumn[\hsize\textwidth\columnwidth\hsize\csname @twocolumnfalse\endcsname

\title{\bf 
Traveling through potential energy landscapes of disordered
materials: the activation-relaxation technique }

\author{Normand Mousseau~\cite{mousadd}}

\address{Computational Physics --- Faculty of Applied Physics,
Technische Universiteit Delft, Lorentzweg 1, 2628 CJ Delft, the Netherlands}

\author{G. T. Barkema~\cite{barkadd}}

\address{HLRZ, Forschungszentrum J\"ulich, D-52425 J\"ulich, Germany}

\maketitle

\begin{abstract}
A detailed description of the activation-relaxation technique (ART) is
presented.  This method defines events in the configurational energy
landscape of disordered materials such as {\it a}-Si, glasses and
polymers, in a two-step process: first, a configuration is activated
from a local minimum to a nearby saddle-point; next, the configuration
is relaxed to a new minimum; this allows for jumps over energy barriers
much higher than what can be reached with standard techniques. Such
events can serve as basic steps in equilibrium and kinetic Monte Carlo schemes.
\end{abstract}
\pacs{PACS: 82.20.Wt, 02.70.Rw, 61.43.Dq and 73.61.Jc.}

\vskip2pc]
\narrowtext

\section{Introduction}

Microscropic structural phenomena often proceed on time scales
remarkably long compared to those of the atomistic oscillations.  This
is the case, for example, for glassy materials where microscopic dynamics
takes place over time scales orders of magnitude larger than that
associated with the natural atomistic time scale, set by the phonon
frequency of typically $10^{13} Hz$.  Such a discrepancy is best
understood from the configurational energy landscape: the system finds
itself in a deep minimum surrounded by energy barriers which are many
times larger than its temperature.  Only rare fluctuations of thermal
energies will allow the system to jump over a barrier and move to a new
minimum.  Typically, the rate for such jumps decreases exponentially
with increasing barrier height, and may reach macroscopic values --- of
the order of seconds or more, rendering the study of these phenomena
rather difficult.

These long time scales are especially prohibitive for numerical
studies. Traditional methods for the study of structural relaxation are
of two kinds: molecular dynamics (MD) and Monte-Carlo (MC). MD is based
on the direct integration of the equations of motion. In order to
ensure the stability of the solution, the integration step cannot be
larger than a fraction of a typical phonon vibration, i.e., somewhere
between 1 and 10 fs. Depending on the number of atoms, the interaction
potential, and the speed of the computer, the total number of steps can
reach $10^4$ to $10^7$, which translates into a time-scale on the order of
nanoseconds; this is still far from the experimental time-scale for
structural relaxation of glassy materials. Because of the nature of MD,
improvements beyond the linear level are particularly difficult to
achieve.  Recently, a promising scheme involving a mixture of
transition-state theory and MD has achieved a significant speed-up in
the simulation of a model system~\cite{voter97}; it is however too
early to say how successful this scheme will be for generic problems.

The inherent limitation to the degree of structural relaxation achieved
with MD does not apply {\it a priori} to MC schemes. Here, the
speed of structural relaxation is mostly determined by the nature of
the attempted moves.  Until now, most algorithms have used moves
defined in real space, involving the displacement of either one or a
limited number of atoms. Single-atom moves are rather efficient in
liquids~\cite{mc}; however, they are not as successful in reproducing
the collective nature of structural relaxation associated with the slow
dynamics of glassy and amorphous materials.  Algorithms with more
complex moves exist: the bond-switching algorithm of Wooten, Winer and
Weaire,\cite{www} for instance, succeeds in producing some of the best
continuous random network models of amorphous semiconductors.  Such
algorithms are however problem-specific, and their dynamics generally
unphysical.

In lattice models like the Ising model, it is often possible to
move from microscopic events --- single spin-flips in the traditional
Metropolis and heat-bath Monte-Carlo simulations -- to collective
events determining the behavior over longer times --- flips of clusters
of spins. Doing so can lead to a substantial improvement in the speed
of algorithms, especially near the critical temperature where the
correlation length and thus the cluster size diverges. The cluster
algorithm of Swendsen and Wang~\cite{swendsen87} for example, can
increase the computational performance of the simulation by many orders
of magnitude compared to single-spin flip algorithms.

In this paper, we give a detailed description of a recently proposed
method which introduces a similar change of paradigm for
continuum-based models: from the microscopic single-atom displacements
to collective moves which form the basis of the activated processes in
glassy and amorphous materials. This method, the activation-relaxation
technique (ART), has already been applied with success to amorphous
semiconductors and metallic glasses\cite{barkema96,mousseau97a,mousseau97b}.
With a similar algorithm, Doye and Wales have studied the
potential energy surface of small Lennard-Jones clusters\cite{doye97a}.

An event in ART is defined as a move from a local energy minimum
$\vec{M}^{(0)} \equiv (\vec{x}^{(0)}_1,\dots,\vec{x}_N^{(0)})$ to another
nearby minimum $\vec{M}^{(1)} \equiv (\vec{x}_1^{(1)},\dots,\vec{x}_N^{(1)})$
following a two-step process mimicking a physical activated processes:
\begin{description} 
\item[i)] the {\bf activation} during which a
configuration is pushed from a local minimum to a nearby saddle-point;

\item[ii)] the {\bf relaxation} which brings the configuration from this
saddle-point to a new local minimum.
\end{description}

By defining the moves in the $3N$-dimensional space controlling the
dynamics of relaxation -- the configurational energy landscape -- ART
removes any constraint on the type of real space moves allowed. This is
particularly important in disordered and complex materials where events
can involve very complex local or collective rearrangements that are
hard to foresee.

This paper is organized as follows: we first present the
activation--relaxation technique.  The following section discusses the
implementation of the algorithm. We finally show examples of events in
amorphous silicon ({\it a}-Si) and silica glas ({\it g}-SiO$_2$). 

\section{The activation--relaxation technique}

In many materials and systems, the dynamics can be accurately described
as a sequence of metastables states separated by energy barriers high
compared to $k_BT$, the typical energy scale at the atomic level. Such
metastable configurations will remain essentially unchanged on a time
long compared with the natural time-scale set by lattice vibrations,
and can be well identified by the atomic positions at zero K, i.e., by
the local minimum of the configurational energy landscape. Knowledge of
the distribution and properties of these local minima is sufficient for
determining the thermodynamical properties of the system. To understand
the dynamical properties of these materials, however, a knowledge of
the rates controlling the jumps from one minimum to another is also
necessary.

To a first approximation, the dynamics in these materials is determined
by the activation energy, i.e. the energy needed to bring a
configuration from the local minimum to a nearby saddle-point. Because
of the exponential nature governing the energy fluctuations, any event
following another trajectory, with by definition an energy higher than
that at the saddle-point, will be much less probable and can be safely
neglected.  \cite{entropy}  For the simplest characterization of the
non-equilibrium properties or dynamics of a disorder material away from
the glass transition it is therefore sufficient to map the continuous
configurational energy landscape onto a network formed by minima
connected via trajectories going through first-order
saddle-points.\cite{saddlepoint} The current ART method provides a
local prescription for exploring this simplified space through hops
from a local minimum to another (events).

By defining the events directly in the configurational energy
landscape, which, as we have seen, fully determines the dynamical and
equilibrium properties of a material, ART becomes much less sensitive
to the details of the real space configuration. Doing so, it refrains
from defining {\it a priori} the type of atomic rearrangements leading
to structural relaxation. In effect, it is the system itself which
determines the appropriate atomic processes, in much closer agreement
with real processes.  Such a change in paradigm, from real to
configurational space, is particularly necessary for the study of
glassy materials where an unambiguous description of real-space
configurations in terms of neighbor lists, coordination defects, etc./
is generally impossible to give.  ART is {\it a priori} blind to the
details of real space configurations; all ART needs is a local and
continuous description of an energy-landscape; discontinuous energy
landscapes, as, for instance, in discrete spin models, cannot be
differentiated and thus forces are not defined. Any continuous
interaction potential however, from Lennard-Jones to LDA, can in
principle be used with ART.

As mentioned in the introduction, the activation-relaxation technique
consists of two parts: a path from a local energy minimum to a nearby
saddle-point --- the activation; and a trajectory from this point to a
new minimum --- the relaxation. 

The relaxation to an energy minimum poses no particular challenge: it is
a well-defined and well-behaved operation for which a number of efficient
algorithms are available (see, for example, Ref. \onlinecite{numrec}). 

The activation from a minimum to a saddle-point requires more care: to
our knowledge, no theoretical framework exists that allows for finding
the complete set of saddle-points around a local minimum.
A number of works have been devoted to the study of finding
the transition states in clusters and low-dimensional systems. Many of the
techniques, however, start with the knowledge of both minimum states
and try to find the path connecting the two.\cite{2points}  It
is a very different problem to try to find a saddle-point
with the knowledge of only one minimum. Most methods can be traced back
to two techniques, the distinguished coordinate
\cite{rothman80,chekmarev94} and the 
eigenvector-following\cite{doye97a,cerjan81,simons83} algorithms.
Although these methods are generic, neither addresses the question
of the generation of a complete set of saddle-points around a given
minimum.

In steepest-descent --- or zero-temperature Langevin dynamics  where the
velocity is proportional to the force --- all trajectories, including
those starting at a saddle-point, lead to a local energy minimum.  A
naive approach to find the trajectory from a minimum to a nearby
saddle-point would therefore be to retrace this path using a
time-reversed zero-temperature Langevin dynamics, or steepest-{\it
ascent} algorithm. This fails, however, since using steepest-ascent
simply corresponds to inverting the sign of the total energy, in effect
exchanging local minima with local maxima. Moreover, the minimum-energy
trajectory leading from a local minimum to a saddle-point is an
unstable trajectory for steepest-ascent; any perturbation sends the
path away from the reversed steepest-descent trajectory.

Within Newtonian mechanics a trajectory from a saddle-point to a
minimum is also time-reversible:  starting at a minimum with 
properly chosen velocities, one would be able to move up to any
saddle-point.  In contrast to time-reversed Langevin dynamics, the
trajectory cannot reach divergent parts of the configurational energy
landscape since the total energy is conserved.  As with time-reversed
zero-temperature Langevin dynamics, though, even a very tiny deviation
can bring the system far away from the saddle-point. Sampling a very
large number of initial random displacements and then targeting the
least divergent trajectories, Dykman {\it et al.} could successfully
find the saddle-points of a chaotic two-dimensional
system.\cite{dykman92} If this approach can work for a simple energy
function in low dimensions, such a hit-and-miss algorithm becomes
hopeless in a large 3N-dimensional space with a computationally expensive
force to evaluate.

At the saddle-point, all eigenvalues of the Hessian but one are positive.
The energy landscape resembles a valley going down along the eigen
directions corresponding to the negative eigenvalue.  Leaving the
saddle-point by steepest-descent we follow the floor of the valley to
eventually arrive at a nearby minimum.  This suggests immediately a
local algorithm which should be more stable than the
steepest-ascent:  to define a trajectory to a saddle-point, the
configuration is moved in such a way as to minimize the force along all
directions but the one corresponding to the lowest eigenvalue. This
eigenvalue is identified with the local bottom of the valley, and
the configuration is moved against the force along this direction. A
small displacement away from the bottom of the valley would be
corrected for by the $(3N-1)$-dimensional minimization, making the
trajectory stable.  Intuitively, this line and the path of
steepest-descent should run mostly parallel; they are not identical
though, and sometimes diverge.

In most circumstances, this algorithm will converge to a saddle-point.
Because we consider here the maximization along a single
eigendirection, this algorithm will not lead to second- or
higher-order saddle-points.  This is in essence what was proposed by
Cerjan and Miller for the location of transition states in
low-dimensional energy surfaces\cite{cerjan81}, and what was used for
an extensive study of a 13-atom LJ cluster by Doye and
Wales\cite{doye97a}.

Because of its $N^3$ requirements, this algorithm becomes rapidly too
computer intensive for realistic bulk systems, often demanding many
hundreds of atoms with a costly energy function.  We must therefore
find another algorithm which does not require evaluation of the full
Hessian matrix at each step.

The current implementation of ART follows a modified force vector
$\vec{G}$, obtained by {\it inverting} the component of the force
{\it parallel} to the displacement from the current position to the
local minimum $\vec{r}=\vec{X}-\vec{M}^{(0)}$
while minimizing all other $3N-1$ directions:
\begin{equation}
\vec{G} = \vec{F} - (1+\alpha) (\vec{F}\cdot\hat{r}) \hat{r}
\label{eq:art}
\end{equation}
where $\hat{r}$ is the normalized vector parallel to $\vec{r}$,
$\vec{F}$ is the total force on the configuration as calculated using
an interaction potential, and $\alpha$ is a control parameter.  This
equation is applied iteratively until the force parallel to the
displacement from the minimum $\vec{F}\cdot\hat{r}$ changes sign from
negative to positive. Generally, the force perpendicular to the
displacement decreases rapidly after a few iterations, bringing the
configuration close to the steepest-ascent trajectory.  For a
steepest-ascent path perfectly parallel to $\vec{r}$, the modified force of
Eq.~(\ref{eq:art}) strictly sticks to the floor of the valley up to
the saddle-point; for steepest-ascent trajectories perpendicular to
$\vec{r}$, the algorithm fails. From experience, such trajectories are
rare and the algorithm generally converges to a saddle-point.

Since moves are defined in the configurational energy landscape,
vectors in Eq.~(\ref{eq:art}) have $3N$ components both for the force
and the position;\cite{volume} the displacement of the configuration
from a local minimum to a nearby saddle-point may therefore involve
{\it any} number of atoms --- from one to all $N$ atoms.

In disordered networks, it is unlikely that the lowest
eigenvalue of the Hessian matrix is degenerate. There are therefore
always only two valleys stemming out of the local minimum,
corresponding to the positive and negative direction of the lowest
eigenvector.  Thus, following valleys from the minimum either along
the lowest eigenvector or the modified force leads
to only two saddle-points, whereas a system typically has many, even
thousands of saddle-points.  Even worse, these two directions
correspond, in bulk materials, to long-wavelength distorsions and do not
lead to interesting events.  Finding a way to avoid these directions
can be a difficult task.

One approach, taken by Doye and Wales for the study of a 13-atom
Lennard-Jones cluster, is to select in turn each of the eigendirections
of the Hessian at the minimum and follow it to a nearby
saddle-point.\cite{doye97a}  Since there are only 78 such directions,
only a fraction of the many ($\approx 10^3$, see Ref.~\onlinecite{wales97})
saddle-points can be reached this way from the minimum; local information
around the minimum is insufficient to locate all valleys leading to
saddle-points. Moreover, the repeated calculation of the Hessian is an
expensive operation for large systems. 

We propose a few approaches that do not require $O(N^3)$ operations
and work for a wide spectrum of circumstances;
these are discussed in section \ref{implementation}.

Once a valley has been found, the situation becomes more
straightforward, and we can use either of the algorithms described
above to follow the valley to the saddle-point.

\section{The implementation}
\label{implementation}

The implementation of the method poses no particular conceptual or
computational problems. The whole code, except for the force and total
energy calculation, contains a few hundred lines at most. Its core
consists of three parts: the escape from the harmonic basin, the
convergence to the saddle-point, and the relaxation to a minimum.

\subsection{Escaping the harmonic basin}

The part of the algorithm that is most sensitive to details of the
system studied is the escape from the harmonic basin; different
approaches might have to be tried to find the most effective one.  In
general, open but stiff materials like amorphous semiconductors have a
very small harmonic basin from which it is easy to escape. More compact
materials --- metallic glasses --- or floppier ones --- silica glasses
--- pose more problems.  To ensure a proper sampling of events, any
method for escaping the harmonic basin that leaves out a significant
fraction of the saddle-points should be avoided.

The simplest way of escaping a harmonic basin is to make a random
displacement away from the minimum, involving a single randomly chosen
atom, a cluster, or all atoms. In our experience, for small systems
they all lead to the same type of events; for larger systems, a global
random displacement tends to induce many spatially separated events
which become difficult to disentangle. We therefore prefer a local
displacement for systems of more than a few hundred atoms.

A random direction generally has a sizable overlap with the
softest elastic modes, and tends to fall back to these easily. We
get better results by taking the escape direction for the initial
displacement along the force induced by a small random displacement; this
procedure is essentially equivalent to applying the Hessian matrix to a
random vector, resulting in a first--order suppression of the softest
elastic modes.

For small systems, where the Hessian can be obtained and diagonalized
in a relatively short time, the softest modes can be removed directly
from a random initial direction, or the initial displacement can be
chosen along a linear combination of the stiffest eigendirections.
This approach is rather computationally involved and cannot be
reasonably carried out for systems with more than 100 or 200 atoms.

Once the initial direction is fixed, it is then followed until the
passage of some threshold, indicating that the harmonic region has been
left. This threshold has to be large enough to ensure that the
trajectory does not fold back onto the softest direction while
remaining insde the basin of attraction. We use a combination
of two conditions for determining the point where the configuration has
left the harmonic region surrounding the initial minimum: when the
force component parallel to the displacement either stops increasing or
when the ratio of this component to the perpendicular component is
smaller then a given fraction, we consider the harmonic region left and
the ART procedure as such begins.

In the algorithm used in Refs. \onlinecite{barkema96},
\onlinecite{mousseau97a} and \onlinecite{mousseau97b}, no clear
distinction was made between leaving the harmonic region and
convergence to the saddle-point; instead, an additional repulsive
harmonic potential was introduced, which is added around the minimum
with a strength $A_{rep}$ and a range $r_c$:  
\begin{equation}
E_{rep} = A_{rep} (|\vec{r}|-r_c)^2.  
\end{equation} 
Although
relatively efficient, this approach modifies the local energy landscape
and introduces an artificial length scale $r_c$ in the problem. To
reduce the impact of this additional length scale, one can re-initialize
$r_c$ and $A_{rep}$ at random before each event.

Currently we prefer to take as initial direction the force after a random
displacement, and follow that direction until we leave the harmonic
region, and then follow $\vec{G}$ as defined in Eq.~(\ref{eq:art}) until
the saddle-point is reached.

\subsection{Convergence to a saddle-point}

Convergence to the saddle-point cannot be achieved using standard
minimization techniques because the modified force $\vec{G}$ as defined in
Eq.~(\ref{eq:art}) is not curl-free, i.e., it cannot be obtained from
the gradient of a scalar.  We therefore have to follow closely
the direction of $\vec{G}$ until we reach the saddle-point, indicated
by a change of sign in the component of the force parallel to $\vec{r}$.
Many simple algorithms can readily be adapted for this purpose.
Making small displacements in the direction of $\vec{G}$ is the
most obvious choice for reaching a saddle-point. Such a crude
method, however, is rather unstable and can easily enter into
oscillations or severe slowing down.

The {\bf conjugate-gradient} (CG) algorithm provides an easy solution
to this restriction by ensuring that the new displacement will be in
a direction conjugate to the previous ones.\cite{numrec}  The line
minimization along a direction $\hat{h}$ required in the CG
implementations of numerical packages, however, are based on the
existence of a total energy - which cannot be defined.  We replace it
by a root-finding algorithm of $\vec{G} \cdot \hat{h}$.  In general,
only a couple of force evaluations are necessary to reach that point.

The {\bf Levenberg-Marquardt} (LM) algorithm~\cite{numrec} proposes a
mixture of steepest-descent and a full-fledged second-order Hessian
minimization technique. Away from the harmonic regime, the
steepest-descent controls the optimization; as steps get smaller and
the space becomes more convoluted, the information contained in the
Hessian matrix starts being used.

If applied directly, the LM algorithm is rather computer intensive and
does not suit our need. However, it is possible to simplify the
algorithm while retaining much of its advantages. We keep here the
steepest-descent part untouched but use a local Hessian, which contains
only the $3\times3$ blocks along the diagonal of the full Hessian,
where both derivatives of the energy are belonging to the same atom.
\cite{steinhardt}
Although this is a rather crude approximation to the real Hessian, it
suffices to reach with a reasonable efficiency the saddle-point and
the minimum on the other side.

At each step, the force and the local Hessian are evaluated. The
displacement is then calculated using a parameter $\lambda$ which is
varied depending on the success of the step:
\begin{equation}
\Delta \vec{X} = \cdot ({\cal H}^{-1}) \vec{F}  
\end{equation}
with
\begin{equation}
{\cal H} = H + \lambda I 
\end{equation}
where $I$ is the identity matrix, and $H$ is the local Hessian.  For
large $\lambda$, the right-hand term dominates, and the algorithm
reduces to a steepest-descent with step-size $\lambda^{-1}$ times the
force.  When the step is too large, $\lambda$ is increased, otherwise
it is decreased.

Both LM and CG require a similar number of steps.  A negative point of
LM is that in order to be computationally efficient, a local Hessian
should be calculated analytically, which is not easy if the force is
taken from already written subroutines or packages.  Therefore, we
tend to prefer CG.

\subsection{Relaxation to a minimum}

Although any method could be used for the relaxation ot the minimum, we
prefer to use the same algorithm as for the convergence to the
saddle-point. In general, it is not necessary to have a very precise
convergence, just a few significant digits (of the order of 0.01 \AA)
suffice. Because of its stability, the convergence to a minimum is
often faster than that to a saddle-point. 

Depending on the material or system, it takes roughly between 100 to
500 force evaluations to converge to a saddle-point, and from 50 to 300
steps to reach an acceptable minimum. For a 500-atom unit cell and a
relaxation of roughly one ART step per atom, this means between 100~000
and 1~000~000 force evaluations.

\section{Example: events in amorphous silicon and silica glass}

To illustrate the real-space working of the algorithm, we present
events created in {\it a}-Si and {\it g}-SiO$_2$. 

A 1000-atom cell of {\it a}-Si was obtained following the prescription
given in Ref.~\onlinecite{barkema96}: starting from a randomly packed
cubic cell, ART\hfill is\hfill applied successively \hfill until

\begin{figure}
\epsfxsize=7cm
\epsfbox{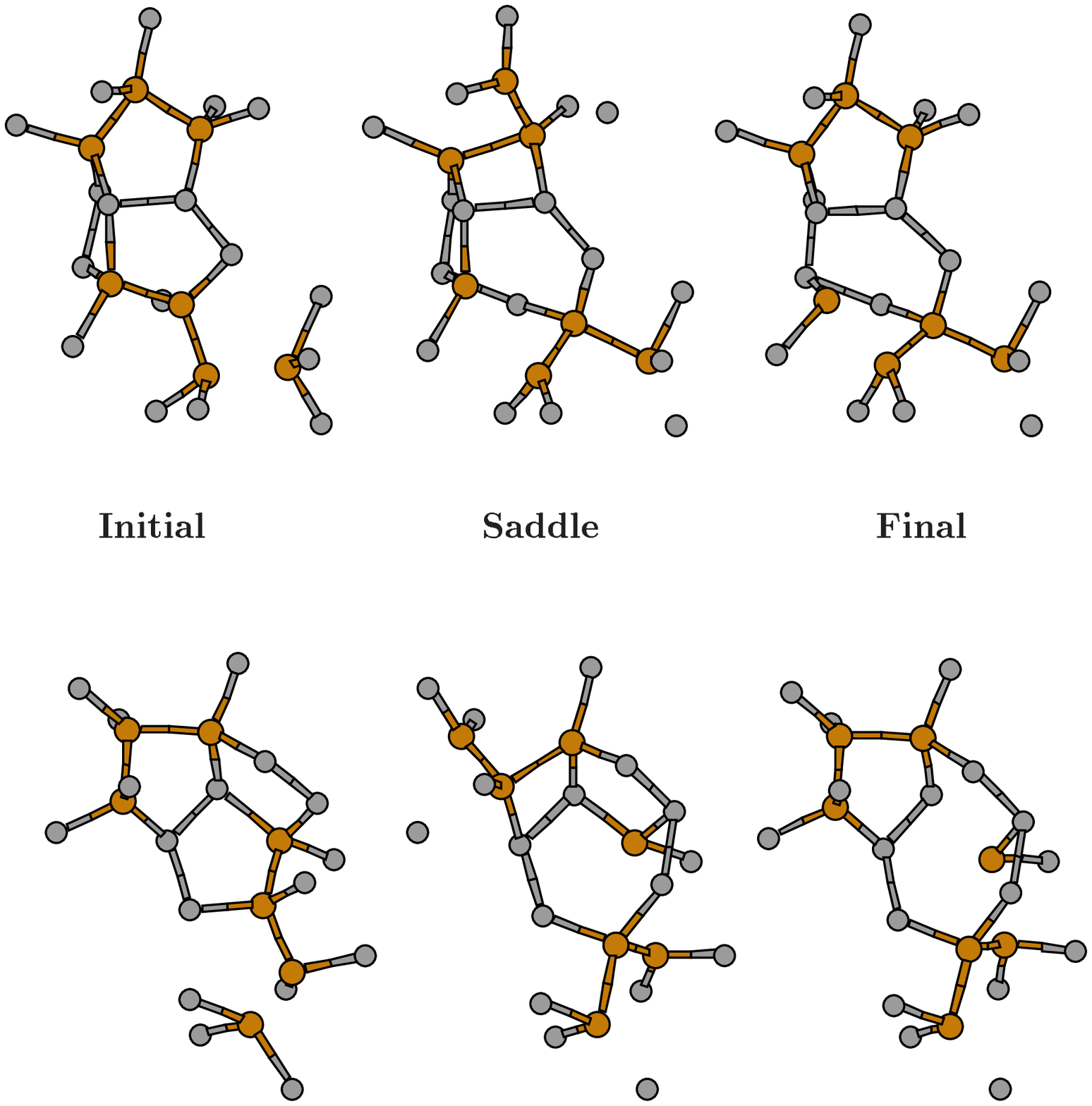}
\vspace{-1.2cm}
\caption{An event in the simulation of amorphous silicon. From left to
right, the initial, saddle-point, and final configurations are shown. The
top and bottom rows correspond to different viewing angles of the same
event. Dark atoms change their bonding environment during the
event; light atoms are nearest-neighbours of the dark atoms.  Activation
energy: 5.74 eV, energy difference from initial to final
configuration: 2.30 eV. 
\label{amorfevent}}
\end{figure}

\noindent the configuration reaches
a stable energy. To obtain a low-energy configuration, we use the standard
Metropolis algorithm where a new configuration is accepted with
probability 1 if the energy is lower than that in the original
configuration, otherwise with probability $\exp(-\Delta E/k_BT)$. The
temperature as such is fictitious and we find that $k_BT=0.25$ gives
satisfactory results.  As in Ref.~\onlinecite{barkema96}, we use a
modified Stillinger-Weber\cite{sw} interaction potential with a
three-body force twice the original value to remove the liquid-like
features of the amorphous phase associated with the original SW.

One event obtained in the relaxed structure is shown in Figure
\ref{amorfevent} from two difference angles. In the bottom
representation, we can see how the configuration passes from three
five-membered rings (initial) to one five- and one eight-membered ring
(final).  In the process, four bonds are broken and four are created,
preserving the total coordination, and the displacement incurred by the
atoms is 2.3 \AA. This event has an activation energy of 5.74 eV and
the final configuration is 2.30 eV higher than the initial one.

\begin{figure}
\epsfxsize=7cm
\epsfbox{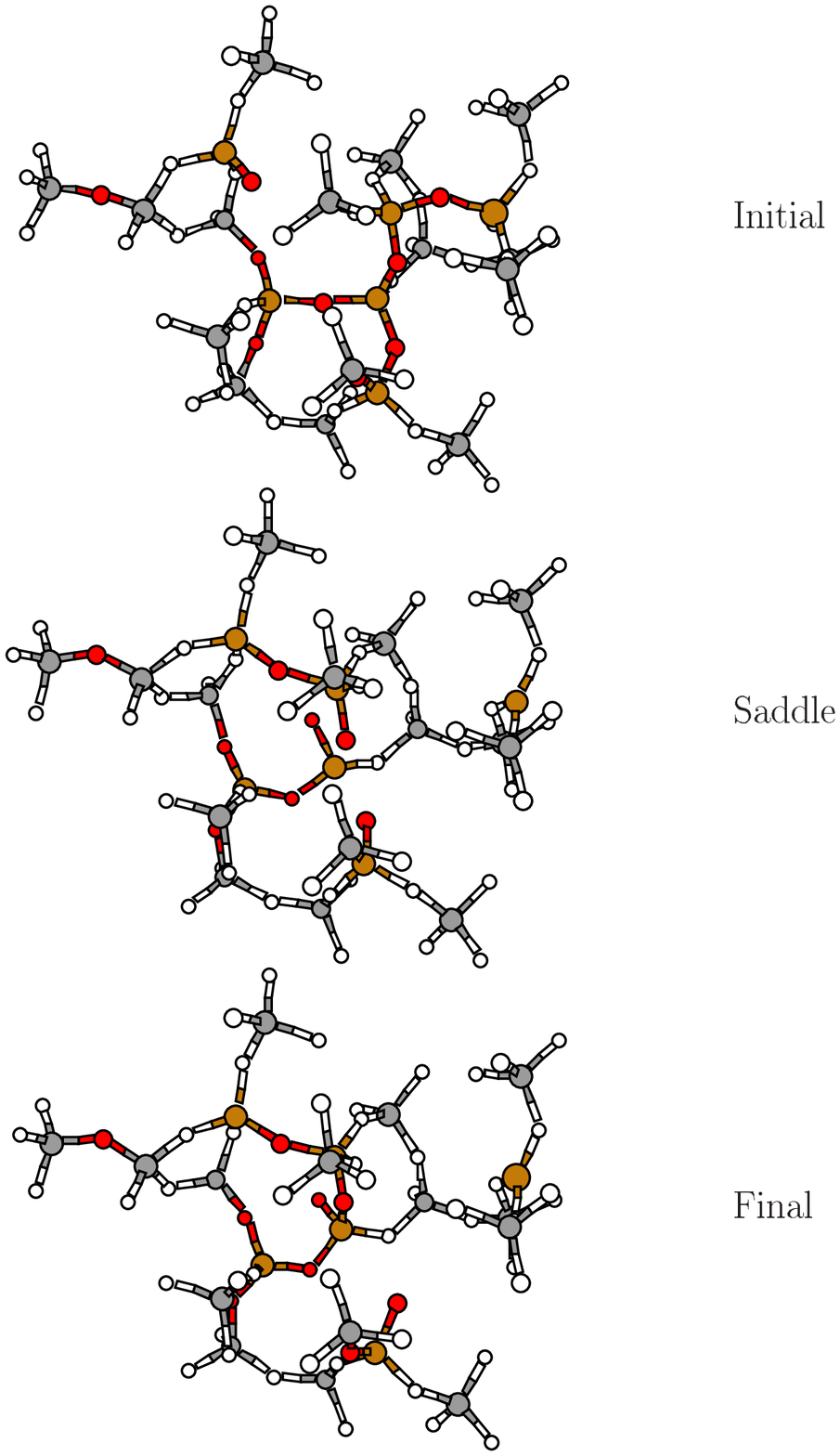}
\caption{An event in the simulation of silica glass. From top to bottom,
the initial, saddle-point, and final configurations are shown. Large
circles represent Si, small ones O atoms. Dark atoms either
change their bonding environment or move by more than 0.75 {\AA} during
the event; light atoms are near-neighbours of the dark atoms. 
Activation energy: 10.84 eV, energy difference from initial to final
configuration: 4.25 eV.
\label{glassevent}}
\end{figure}

For silica glass, we use a 576-atom configuration relaxed from the melt
using molecular dynamics \cite{beckers97}. The initial relaxation was
done using the full Vashishta {\it et al.} potential\cite{vashishta90}
while ART was applied using a screened version of the same
potential.\cite{nakano94}  Figure \ref{glassevent} shows an event in
this structure. Because of \hfill its more  \hfill open \hfill nature, events 

\noindent in silica tend
to involve more atoms than in amorphous silicon. Total atomic
displacement between initial and final configurations is 6.8 \AA{} with
three broken and two created bonds and many tens of atoms involved at a
lower degree during the activation and relaxation phases. The
activation energy is considerable, at 10.84 eV, with the new
configuration 4.25 eV higher in energy than the initial one.

The characterization of events both in {\it a}-Si and {\it g}-SiO$_2$
is difficult:  although each event normally involves less than 10 to
12 bonds being broken or created, many more atoms can move
significantly, rendering visualization complicated. We are currently
working on a systematic study of events in both materials.  

\section{Conclusion}

By defining events directly in the configurational energy landscape,
the activation-relaxation technique provides a generic approach to
study relaxation in complex systems such as glassy and amorphous
materials, polymers, and clusters.  Real space moves are determined by
the system itself and represent the most likely physical trajectories
followed during relaxation. ART is much less sensitive to the
slowing down caused by increasing activation energy barriers than
standard MC and MD approaches.

Already ART has produced results which could not be achieved via other
techniques: it has produced well-relaxed samples of
{\it a}-Si,\cite{barkema96}
{\it a}-GaAs,\cite{mousseau97a,mousseau97b} Ni$_{80}$P$_{20}$,
\cite{barkema96} and
minimum-energy configurations of clusters of Lennard-Jones
particles.\cite{doye97a}  The examples of events presented here
demonstrate that ART can easily reach regions of the energy landscape
which are difficult to sample using more standard techniques.
This paper provides the necessary description of the algorithm to
allow for a rapid application of ART to a wide range of problems.

\section{Acknowledgements}

We acknowledge useful and interesting discussions about ART with J. P. K.  
Doye, M. I. Dykman, S. W.  de Leeuw, and V. Smelyanski, and partial
support by the Stichting FOM (Fundamenteel Onderzoek der Materie) under
the MPR program.

\bibliographystyle{prsty}

\begin{thebibliography}{99}
\bibitem{volume}
Besides the atomic coordinates, it is also possible to include
additional
degrees of freedom in the vector on which ART is applied. For some
very dense materials, e.g., metallic glasses, relaxation may be
improved significantly by allowing the volume to vary also.
\end{thebibliography}

\end{document}